\documentclass{ifacconf}
\usepackage[fixamsmath]{mathtools}
\mathtoolsset{showonlyrefs,mathic,centercolon}
\usepackage{mathrsfs,amsfonts} 
\usepackage{amsmath}			
\usepackage{amssymb}			
\usepackage{mathcomp}	

\usepackage{graphicx}    
\usepackage{natbib}        
\usepackage{xcolor}
\usepackage{wrapfig}

\usepackage{algorithm}
\usepackage{algorithmicx}
\usepackage{algpseudocode}

\newcommand{\E}{\mathbb{E}}
\newcommand{\R}{\mathbb{R}}
\newcommand{\N}{\mathbb{N}}

\begin{document}
\begin{frontmatter}

\title{Deep reinforcement learning for scheduling in large-scale networked control systems} 

\thanks[footnoteinfo]{Research supported
	by the German Research Foundation (DFG) - 315248657. \\
8th IFAC Workshop on Distributed Estimation and Control in Networked Systems NECSYS 2019: Chicago, USA, 16{-}17 Sep. 2019}

\author[First]{Adrian Redder} 
\author[First]{Arunselvan Ramaswamy} 
\author[First]{Daniel E. Quevedo}

\address[First]{Faculty of Computer Science, Electrical Engineering and Mathematics, Paderborn
	University (e-mails: aredder@mail.upb.de, arunr@mail.upb.de, dquevedo@ieee.org)}

\begin{abstract}                
	This work considers the problem of control and resource allocation in networked systems. To this end, we present DIRA a Deep reinforcement learning based Iterative Resource Allocation algorithm, which is \textit{scalable} and \textit{control-aware}. Our algorithm is tailored towards large-scale problems where control and scheduling need to act jointly to optimize performance. DIRA can be used to schedule general time-domain optimization based controllers. In the present work, we focus on control designs based on suitably adapted linear quadratic regulators. We apply our algorithm to networked systems with correlated fading communication channels. Our simulations show that DIRA scales well to large scheduling problems.
\end{abstract}

\end{frontmatter}

\section{Introduction}\label{sec:intro}
Sequential decision making in uncertain environments is a fundamental problem in the current data-driven society.
They occur in cyber-phyical systems such as smart-grids, vehicular traffic networks, Internet of Things or networked control systems (NCS), see\textit{,} e.g., \cite{Iot-large-scale}.
Many of these problems are characterized by decision making under resource constraints.
NCS consist of many inter-connected heterogeneous entities (controllers, sensors, actuators, etc.) that share resources such as communication \& computation.
The availability of these resources do not typically scale well with system size; hence effective resource allocation (scheduling) is necessary to optimize system performance.


A central problem in NCS is to schedule data transmissions to available communication links. Traditionally, this is tackled by periodic scheduling or event-triggered control algorithms, see \cite{Heemels_triggered} and \cite{Park_periodic}. To solve the sensor scheduling problem, \cite{ramesh2013design} proposed a suboptimal approach, where scheduler and control designs are decoupled. They also show that for linear single-system problems with perfect communication, it is computationally difficult to find optimal solutions. It must be noted that scheduling controller-actuator signals is a non-convex optimization problem, see \cite{peters2016controller}.

Networked controllers often use existing resource allocation schemes, however, such schemes typically reduce waiting times, and/ or maximize throughput, see \cite{sharma2006complexity}. However, such approaches are not context aware, i.e. they do not take into account consequences of scheduling decisions on the system to be controlled. An additional challenge for scheduling problems stems from the inherent uncertainty in many NCS. Specifically, an accurate communication dynamics model is usually unknown, see also \cite{eisen2018learning}. To this end, a combined scheduling and control design, which can use system state information and performance feedback, where optimal control and resource allocation go hand in hand to jointly optimize performance is highly desirable. \textit{In the present work, we tackle this problem by combining deep reinforcement learning and control theory.}

\begin{figure}[t]
	\centering
	\includegraphics[width=1.00\linewidth]{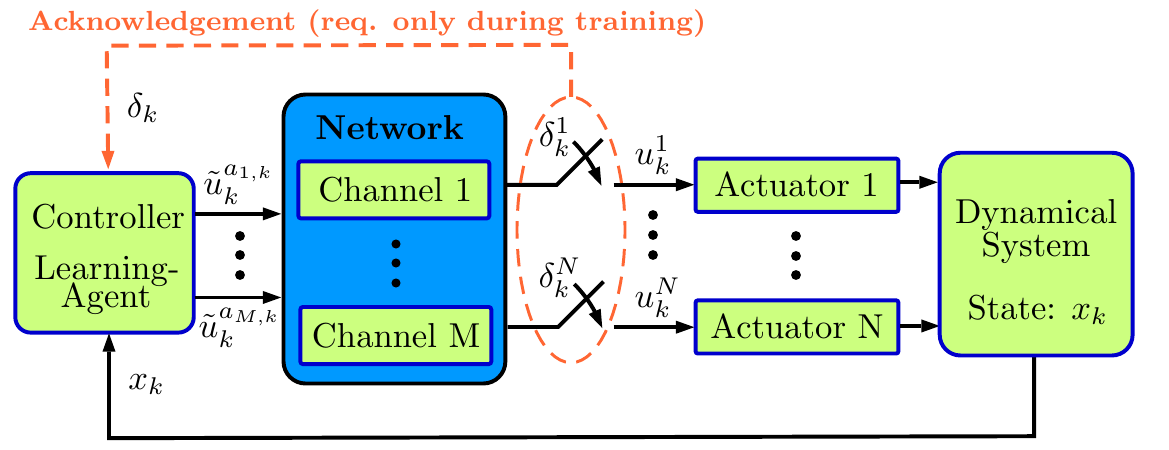}
	\caption{Networked system with $N$ independent actuator dynamics, controlled via $M$ communication channels. At every time-step the controller uses state information to allocate actuator inputs to each channel. }
	\label{fig: ncs}
\end{figure}



Deep reinforcement learning (RL) is a combination of RL and neural network based function approximators that tackles Bellmans curse of dimensionality, see \cite{bellman1957dynamic}. 
The most popular algorithm is deep Q-learning, which achieved (super) human level performance in playing Atari games, see \cite{mnih2015human}. Deep RL has been applied successfully to various control applications. \cite{baumann2018deep} applied the recent success of actor-critic algorithms in an event-triggered control scenario. In \cite{lenz2015deepmpc}, the authors combined system identification based on deep learning with model predictive control. 

Deep RL has been applied to resource allocation problems, see for example \cite{mao2016resource}. \cite{demirel2018deepcas} and \cite{leong2018deep} have shown the potential of deep RL for control and scheduling in NCS. However, these solutions do not extend well to large-scale problems, 
since the combinatorial complexity of the proposed algorithms grows rapidly with the number of systems and available resources.

The main contribution of this work is DIRA, a Deep RL based Iterative Resource Allocation algorithm. This algorithm is tailored towards large-scale resource allocation problems with a goal to improve performance in a control-aware manner. The algorithm uses system state information and control cost evaluations (performance feedback) to improve upon an initial random scheduling policy. 
Further, DIRA has the ability to adapt to a given control policy which allows for such performance feedback. For control, we present a simple yet effective design based on time-varying linear quadratic regulation. DIRA and the controller act jointly to optimize control performance. Our simulations show that DIRA scales well to large decision spaces using simple neural network architectures. 
Finally, we would like to highlight that DIRA does not require a network model, but implicitly learns network parameters. \textit{To the best of our knowledge DIRA is the first scalable control-aware scheduling algorithm that accommodates correlated fading channels with unknown parameters.}



\section{Scheduling and Control}
\subsection{Networked control system architecture}\label{sec: NCS-architecture}
We consider a large networked control system with $N$ discrete-time linear subsystems with state vectors  $x^{i}_k \in \mathbb{R}^{n_i}$, control inputs $u^{i}_k \in \mathbb{R}^{m_i}$ and i.i.d. noise processes $w_k^i \sim \mathcal{N}(0,\Sigma_{w^i})$. Here, $k \ge 0$ denotes the discrete time index and $1\le i \le N$ denotes the subsystem index. The reader is referred to Fig. \ref{fig: ncs} for an illustration. We define concatenated state, control and noise vectors by $x_k \coloneqq (x_k^i)_{1\le i \le N}$, $u_k \coloneqq (u_k^i)_{1\le i \le N}$ and $w_k \coloneqq (w_k^i)_{1\le i \le N}$, respectively.  Let $n := \sum_{i=1}^{N}n_i$ and $m := \sum_{i=1}^{N}M_i$. 

In this paper we allow for coupled subsystems. However, the input (actuator) dynamics are assumed to be independent. Hence the overall linear system dynamic is given by:
\begin{equation}\label{eq: system}
	\begin{split}
		x_{k+1} &= A x_k + \underbrace{\begin{pmatrix} B^{1}& & 0 \\ & \ddots & \\ 0& & B^{N} \end{pmatrix}}_{\coloneqq B} u_k + w_k.
	\end{split}
\end{equation}
We define the single state quadratic costs $g(x,u) \coloneqq x^\top W x + u^\top R u$, where $W$ is a positive semi-definite matrix and $R$ is a positive definite matrix.
We assume that the pair $[A,B]$ is controllable and that the pair $[A,W^{1/2}]$ is observable.

The system is controlled by a central controller which has access to all state vectors $x_k^i$. The key feature of the problem at hand is that candidate control inputs $\tilde{u}_k^i$ need to be transmitted over a limited number of fading channels prone to dropouts. Specifically, we assume that at every time-step the controller can access $M$ independent communication channels. Hence resource allocation of $M$ channels to $N$ controller-actuator links is necessary. In summary, the controller needs to select pairs $(\tilde{u}^{a_{j,k}}_k, j)$, where $1 \le j \le M$ and $a_{j,k} \in \{1,\ldots,N\}$, such that $(\tilde{u}^{a_{j,k}}_i, j)$ denotes that candidate control signal $\tilde{u}^{a_{j,k}}_i$ is sent to subsystem $a_{j,k}$ using channel $j$ at time $k$.

\begin{rem}
	In this paper we assume that the probability of success for a particular channel is independent of other channel usage.
	Hence, using multiple channels in parallel enhances the probability of successful transmission.	
	This allows for general decision spaces, where the controller can transmit candidate control signals $\tilde{u}_k^i$ using multiply channels.  Further,  scenarios where $M \ge N$ are also included. The reader must note, that our scheduling-algorithm does not rely on the above independence assumption, see section \ref{sec:algorithm}. Specifically, it always strives to improve control performance.
\end{rem}


With regards to communication, we use correlated fading channels as described in \cite{FiniteMarkovChain} to model the communication network. 
More precisely, we describe $M$ channels as Markov processes $\{Z_k^j\}$ with state-space $\mathcal{Z}_j = \left\{z^j_0,z^j_1, \ldots, z^j_{K-1} \right\}$, parameterized by tuples 
\begin{equation}\label{eq: markov_channels}
(T_j,p_j,e_j) \quad \forall j\in\{1,\ldots,M\}, 
\end{equation}
where $T$ denotes the transition probability matrix, $p$ denotes the steady state probability vector and $e$ denotes the drop-out probability vector. In each channel state $z^j_d$, fading results in a communication drop out with a probability according to the $d$-th component of $e_j$.We assume, that at any time $k$ the controller receives acknowledgment from each actuator for successful receptions of $\tilde{u}^i_k$ during an initial ``training phase'', see section \ref{sec:algorithm}. Let $\delta^i_k$ be random variables such that $\delta_k^i = 1$, if a control signal $\tilde{u}^i_k$ is successfully received at actuator $i$. We define the NCS inputs as 
\begin{equation}
u_k^i \coloneqq 
\begin{cases}
\tilde{u}_k^{i}, & \text{if } \delta_k^i = 1 , \\
0,& \text{otherwise}.
\end{cases}
\end{equation}
This corresponds to a zero-input strategy in case of no available control data. We refer the reader to \cite{Schenato2009} for a comparison between zero-input and hold-input strategies over lossy networks.



We consider that the parameters defining the communication network \eqref{eq: markov_channels} are unknown. 
Estimating unknown parameters for a possibly time varying environment in an online manner is a difficult problem, \cite{eisen2018learning}. Often, the estimation relies on repeated test signals, with a large enough sample size, which could be expensive.
In this work, we assume that the controller has to act solely based on information gathered when transmitting over the network.


\subsection{Joint scheduling and control problem}
In the described NCS setup, the controller has to schedule each of the $M$ channels to one of the $N$ subsystem actuators. For each channel, $a_{j,k} \in \{1,\ldots,N\}$ defines a decision variable such that $a_{j,k} = i$, if channel $j$ is scheduled to subsystem $i$. 
Hence, at any time $k$ the decision (action) space of the scheduling problem is given by $$\mathcal{A}_s = \left\{ a_k = (a_{1,k}, \ldots, a_{M,k}) \bigm\vert 1\le a_{j,k} \le N, 1\le j \le M \right\}.$$ 

Therefore, the action space has a size of $|\mathcal{A}_s| = N^M$. 
Recall that we allow the controller to use multiple distinguishable resources to close the controller-actuator links.

The controller wishes to find a stationary joint control-scheduling policy $\pi$ mapping states to admissible control-scheduling decisions, i.e. at any time $k$ the controller needs to select $a_k \in \mathcal{A}_s$ and $u_k^i \in \R^{m_i} $ for all $ i \in a_k$. Define the corresponding action space as 
$$\mathcal{A} = \{\{(u_k^{a_{1,k}},1), \ldots, (u_k^{a_{M,k}},M)\} \mid a_{j,k} \in \N_{<N} ,u_k^i \in \R^{m_i} \}. $$ We can represent the joint control-scheduling policy by a pair $(\pi_c,\pi_s)$,
where 
\begin{equation}
\pi_s: \R^n \rightarrow \mathcal{A}_s, \qquad \pi_c: \R^n \times \mathcal{A}_s \rightarrow \mathcal{A}.
\end{equation}

The expected average cost following a stationary policy $\pi$ with initial state $x$ reads
\begin{equation}
\hspace{-0.3cm} J^\pi(x) =  \limsup\limits_{T \rightarrow \infty} \frac{1}{T} 
\underset{\underset{\tilde{u}_k \sim \pi(x_k)}{x_k,u_k \sim \mathcal{E}}}{\E}
\left\{\sum_{k=n}^{T+n-1} \hspace{-0.2cm} g(x_k,u_k) \bigm\vert x_n = x\right\},
\end{equation}
where $\E_{\xi\sim \zeta} \left\{ \cdot \right\}$ denotes the expected value, with $\xi$ distributed according to $\zeta$. Here, $\mathcal{E}$ denotes the system environment represented by the stochastic processes $$(w_k^1,\ldots,w_k^N,Z_k^1,\ldots,Z_k^M, k \ge 0).$$
The direct minimization of $J^\pi(x)$ over all admissible policies for all states $x\in \R^n$ is a difficult problem. This stems from the fact that the space of admissible control signals is discrete-continuous and non-convex. Additionally, the expectation in the above equation is with respect to the Markov process dynamics which are unknown.
All these challenges motivate the use of model-free learning techniques in combination with linear control theory to find a possibly suboptimal solution for the joint scheduling and control problem. 

\section{Deep RL for control-aware resource scheduling }\label{sec: DRL for scheduling}
To obtain a tractable solution, we shall decompose the joint scheduling and control problem into the following parts:
\begin{enumerate}
	\item[(i)] A deep RL based scheduler DIRA, which iteratively picks actions $a_k \in \mathcal{A}_s$ at every time $k$.
	\item[(ii)] A time-varying linear quadratic controller which computes candidate control signals $\tilde{u}_k^i$ based on the scheduling decisions $a_k$ and the approximated success probability of each controller-actuator link.
\end{enumerate} 

We describe the scheduler-design and controller-design in Section \ref{scheduler} and Section \ref{sec: controller}, respectively. In section \ref{sec:algorithm}, we combine these components to obtain an algorithm for joint control and communication.
Before proceeding, we give some background on deep reinforcement learning. 

\subsection{Background on deep RL}\label{sec: DRL}
In RL an agent seeks to find a solution to a Markov decision process (MDP) with state space $\mathcal{S}$, action space $\mathcal{A}$, transition dynamics $P$, reward function $r(s,a)$ and discount factor $\gamma \in (0,1]$. It does so by interacting with an environment $\mathcal{E}$ via a constantly evolving policy $\pi: \mathcal{S} \rightarrow \mathcal{A}$. An optimal solution to an MDP is obtained by solving the Bellman equation given by
\begin{equation}\label{eq: Bellman}
Q^*(s,a) = \underset{s',r \sim \mathcal{E}}{\E} \left\{r(s,a) + \gamma \max_{a' \in \mathcal{A}} Q^*(s',a') \bigm\vert s,a \right\}.
\end{equation}
Deep RL approaches such as deep Q-Learning seek to estimate $Q^*$ (also known as Q-factors) for every state-action pair. The resulting optimal policy $\pi^*$ is given by 
\begin{equation}\label{eq: greedy}
	\pi^*(s) = \arg\max_{a \in \mathcal{A}} Q^*(s,a), \ \forall s\in \mathcal{S}.
\end{equation}
Deep Q-Learning seeks to find the best 
neural network based approximation $Q(s,a; \theta^*)$ of  $Q^*(s,a)$ for every state-action pair, see \cite{mnih2015human}. This is done by performing mini-batch gradient descent steps to minimize the squared Bellman loss  $\left( y_k - Q(s,a,;\theta)  \right)^2$, at every time $k$, with the target values $$y_k = r(s,a) + \gamma \max_{a' \in \mathcal{A}} Q(s',a'; \theta).$$ These mini-batches are sampled from an experience replay $\mathcal{R}$ to reduce the bias towards recent interactions. 

Value function based methods such as Q-Learning seek to find an optimal policy implicitly. On the other hand, it is also possible to directly parameterize a policy and train it to optimize a performance criterion. Examples include actor-critic style algorithms. We refer the reader to \cite{sutton2018introduction} for details on policy based methods.

\subsection{DQN for resource allocation}\label{seq: DQN scheduler} 
Let us say that we are given a control policy $\pi_c$. In principle, we can find a control-aware scheduling policy $\pi_s$ using the DQN paradigm as described in Section \ref{sec: DRL}. For this, we define an MDP $\mathcal{M}_1$, with state space $\mathcal{S}_1 = \R^n$, action space $$\mathcal{A}_1 = \left\{ (a_{1,k}, \ldots, a_{M,k}) \bigm\vert 1\le a_{i,k} \le N, 1\le i \le M \right\},$$ reward signal $r_k = -g(x_k,u_k)$ and discount factor $\gamma \in (0,1]$. 
By assining the negative one-stage cost as a reward, a solution to $\mathcal{M}_1$ minimizes the discounted cost $$\limsup\limits_{T \rightarrow \infty} 
\underset{x_k,u_k \sim \mathcal{E}}{\E} \left\{ \sum_{k=0}^{T-1} \gamma^k g(x_k,u_k) \bigm\vert \pi_c \right\}.$$
Unfortunately, the  direct  application  of  Deep Q-Learning to solve this MDP is infeasible when $N$ and $M$ are large, since the algorithm is usually divergent for large action spaces of size $|\mathcal{A}| = N^M$.
The next subsection presents a reformulation of $M_1$, which goes beyond DQN to address this scalability issue. In section \ref{sec: controller} we present a control policy design $\pi_c$, which adapts to the learned schedule $\pi_s$. 

\subsection{An MDP for iterative resource allocation}\label{scheduler}

The intractability of the scheduling problem for large action spaces, is addressed by exploiting the inherent iterative structure of the resource allocation problem. Let us say that the system state is $x_k$ at time $k$. The scheduler DIRA iteratively picks component-actions $a_{1,k},\ldots,a_{M,k}$ to obtain scheduling action $a_k$. Recall that $a_{j,k} = i$ when subsystem $i$ is allocated to channel $j$. Once $a_k$ is picked, the controller transmits the associated control signals, see Fig. \ref{fig: ncs}. After that, the scheduler receives acknowledgments $\delta_k^i$, which enables the computation of the reward signal $r_k = -g(x_k,u_k)$ (performance feedback).

\begin{figure}[t]
	\centering
	\includegraphics[width=0.85\linewidth]{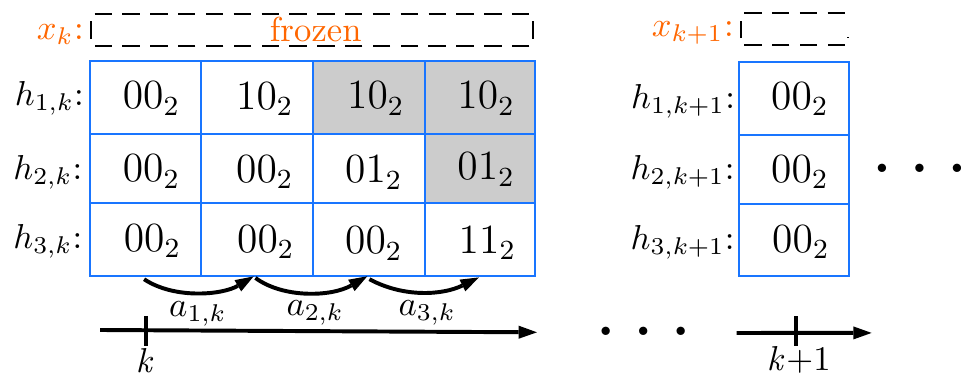}
	\vspace{-0.2cm}
	\caption{Iterative selection procedure for $N=M=3$. The system is in state $x_k$ and $h_{k}$ is initialized to $\underline{0}$. Component actions $a_{j,k}$ are selected as a function of the intermediate states $(x_k,h_k)$. Let us say $a_{1,k} = 2$, $a_{2,k} = 1$ and $a_{3,k} = 3$, then $h_k$ is updated to $(10_2, 00_2, 00_2)$, followed by updates to $(10_2, 01_2, 00_2)$ and $(10_2, 01_2, 11_2)$. Thus $a_k$ = $(2,1,3)$ is selected.}
	\label{fig: selection}
\end{figure}

Let us construct an $M$-dimensional representation vector $$h_{k} := (h_{1,k}, \ldots, h_{M,k} ) \in \mathcal{H},$$ where $h_{j,k}$ is the binary representation of action $a_{j,k}$ and $\mathcal{H}$ denotes the space of all possible representation vectors $h_k$. We define intermediate states as $(x_k,h_k)$, where each element of $h_k$ is assigned iteratively at a fast rate between successive time-steps $k$ and $k+1$, as illustrated in Fig. \ref{fig: selection}.
 
Let $x, x' \in \mathcal{S}_1$ and $h,h' \in \mathcal{H}$.
We say that $(x,h)$ and $(x', h')$ are ``equivalent" iff $x = x'$. We define an equivalence class by $[x] \coloneqq \{ (x',h) \bigm\vert (x',h) \in \mathcal{S}_1\times \mathcal{H},\  x' = x \}$.
Between times $k$ and $k+1$, the system state $x_k$ is frozen, while the representation vector $h_k$ changes.\footnote{Note that other encoding schemes such as one-hot encoding may be used to represent the $a_{j,k}$'s.} Hence all the intermediate states are equivalent. 
The idea of freezing a portion of the state space is inspired by \cite{mao2016resource}. We defined this equivalence, since we assign the same reward $r_k$ to all intermediate actions $a_{j,k}$. This is because the $a_{j,k}$'s are combined to obtain the action $a_k$, which in turn results in one single stage cost.

We therefore have a \textit{natural MDP reformulation} $\mathcal{M}_2$ to embed the above iterative procedure:
\begin{itemize}
	\item[$\mathcal{S}_2:$] contains all state equivalence classes as defined above.
	\item[$\mathcal{A}_2:$] is given by $\{1,\ldots,N\} $.
	\item[$r:$] is given by $-g(x_k,u_k)$.
	\item[$\gamma:$] is the discount factor such that $\gamma \in (0,1]$.
\end{itemize}

In the next section we will solve $\mathcal{M}_2$ using the $DQN$ paradigm. An important consequence of the reformulation is that $\mathcal{M}_2$ has an action space of size $|\mathcal{A}_2| = N$, opposed to $|\mathcal{A}_1| = N^M$.

\section{Joint control and communication}\label{sec: joint}
\subsection{Controller design}\label{sec: controller}
Until now we have considered how to schedule resources for a given control policy. On the one hand, we achieve ``control-awareness" of our scheduler by providing the negative one stage costs as rewards in our MDP, see Section \ref{scheduler}. On the other hand, we achieve ``schedule awareness" by parameterizing a linear quadratic regulator by the expected rates at which the control loops are closed. Specifically, we approximate the success probabilities of each controller-actuator link by a moving average and update the controller during runtime.

The system dynamics in \eqref{eq: system} can be written in terms of the success signals $\delta_k^i$ by defining $\Delta_k = \text{diag}( \left\{I_{m_i}\delta_k^i \right\}_{i=1}^N)$, where $I_{m_i}$ denotes the identity matrix of dimension $m \times m$. Then define $B_{\Delta_k} \coloneqq B \Delta_k$. 

Consider the LQR problem with finite horizon $T$:
\begin{equation}\label{eq: finite_lqr}
\begin{split}
\underset{\{u_k\}}{\min} &  \underset{\underset{k=0,1,\ldots,T-1}{w_k,B_k}}{\E} \left\{ x_T^\top W x_T + \sum_{k=0}^{T-1} x_k^\top W x_k + u_k^\top R u_k \right\}, \\
\text{s.t.} & \qquad x_{k+1} = Ax_k + B_{\Delta_k} u_k + w_k.
\end{split}
\end{equation}
Assume that $B_{\Delta_k}$'s are independent with finite second moments. Then the dynamic programming framework yields the following optimal finite horizon solution to \eqref{eq: finite_lqr}
\begin{equation}
u_k^* = - \left( R + \E \left\{B_{\Delta_k}^\top K_{k+1} B_{\Delta_k} \right\} \right)^{-1} \E \left\{B_{\Delta_k}^\top\right\}K_{k+1}A x_k
\end{equation}
with $K_T = W$, 
\begin{equation}\label{eq: recursive riccati}
\begin{split}
K_k &= A^\top K_{k+1}A +W - A^\top K_{k+1} \E \left\{B_{\Delta_k}\right\} \\ &\times\left( R + \E \left\{B_{\Delta_k}^\top K_{k+1} B_{\Delta_k} \right\} \right)^{-1} \E \left\{B_{\Delta_k}^\top\right\}K_{k+1}A. 
\end{split}
\end{equation}
We will use the steady state controller 
\begin{equation}\label{eq: steady state controller}
u_k = - \left( R + \E \left\{B_{\Delta_k}^\top K_\infty B_{\Delta_k} \right\} \right)^{-1} \E \left\{B_{\Delta_k}^\top\right\}K_\infty A x_k
\end{equation}
\begin{equation}\label{eq: riccati steady state}
\begin{split}
\hspace{-0.3cm}\text{with }K_\infty &= A^\top K_\infty A + W - A^\top K_\infty \E \left\{B_{\Delta_k}\right\} \\ &\times\left( R + \E \left\{B_{\Delta_k}^\top K_\infty B_{\Delta_k} \right\} \right)^{-1} \E \left\{B_{\Delta_k}^\top\right\}K_\infty A.
\end{split}
\end{equation}
It is important to point out that equation \eqref{eq: riccati steady state} may not necessarily have a solution, i.e., \eqref{eq: recursive riccati} may not converge to a stationary value, see Section 3.1. of \cite{BertsekasVI}.
The following lemma establishes conditions such that \eqref{eq: riccati steady state} has a steady state solution. It extends \cite{ku1977further} to the  case where $B$ is disturbed by a multiplicative diagonal matrix. 
\begin{lem}\label{lem: 1}
	A steady state solution for \eqref{eq: recursive riccati} exists if 
	\begin{equation}
		\lambda_{\max}(\Gamma A) < 1, 
	\end{equation}
	where $\Gamma$ is defined by
	{
	$$\Gamma = \text{diag} \left( \left\{I_{n_i} \left(\sqrt{1-\E\{\delta^i_k\}} \right)\right\}_{i=1}^N \right)$$}
	\normalsize
	and $\lambda_{\max}(\cdot)$ denotes the largest absolute eigenvalue.
\end{lem}
\begin{pf}
	Consider the recursive Riccati equation \eqref{eq: recursive riccati}. Define $\alpha = \E\{\Delta_k\}$. 
	Observe that, since $B^\top K_{k+1} B$ is a symmetric matrix it commutes with the diagonal matrix $\Delta_k$. Thus
	$\E \left\{B_{\Delta_k}^\top K_{k+1} B_{\Delta_k} \right\} = \alpha^2 B^\top K_{k+1} B$, since $\delta_k^i$ are i.i.d. Bernoulli.
	Using the above observations we can rewrite \\
	$\E \left\{B_{\Delta_k}\right\} \left( R + \E \left\{B_{\Delta_k}^\top K_{k+1} B_{\Delta_k} \right\} \right)^{-1} \E \left\{B_{\Delta_k}^\top\right\}$
	from \eqref{eq: recursive riccati} as \\ $ \alpha \left(\frac{1}{\alpha^2} R + B^\top K_{k+1} B \right)^{-1} $. 
	
	Notice that
	$B \alpha = \beta B $, with $\beta = \text{diag}( \left\{I_{n_i} \E\{\delta^i_k\}\right\}_{i=1}^N )$.
	Then, the Riccati equation \eqref{eq: recursive riccati} can be rewritten as \\
	$K_k = A^\top(1-\beta)K_{k+1}A + Q + A^\top \beta M_k A$, where \\
	$M_k \coloneqq K_{k+1} - K_{k+1} B \left( \frac{1}{\alpha^2} R +B^\top K_{k+1} B\right)^{-1} B^\top K_{k+1}$
	(as $\beta$ commutes with $K_{k+1}$).
	Now, using arguments that are similar to \cite{ku1977further} we obtain $M_k \le L \ \forall k$. If we define $W' \coloneqq W - A^\top (1-\beta) L A$, then we have
	\begin{equation}\label{riccati bound}
	\begin{split}
	K_k  
	&\le A^\top (1-\beta)^{1/2}K_{k+1} (1-\beta)^{1/2} A + W'.
	\end{split}
	\end{equation}
	Finally, if the eigenvalues of $(1-\beta)^{1/2}A \coloneqq \Gamma A$ lie in the unit circle, then the recursion associated with \eqref{riccati bound} converges by Lyapunov stability theory and so does recursion \eqref{eq: recursive riccati}.
\end{pf}

Under the conditions of Lemma \ref{lem: 1} we will use \eqref{eq: steady state controller} as a time-varying control policy in combination with our iterative scheduling algorithm. Specifically, we calculate $K_\infty$ using a sample based approximation of $\E\{\delta_k^i\}$. In doing this, the controller varies according to the expected rate at which the controller-actuator links are closed and therefore adapts to the scheduling policy $\pi_s$. After a scheduling action $a_k \in \mathcal{A}_s$ is chosen, we transmit
\begin{equation}\label{eq: control signals}
	\tilde{u}_k = - \left( R + \E \left\{B_{\Delta_k}^\top K_\infty B_{\Delta_k} | a_k \right\} \right)^{-1} \E \left\{B_{\Delta_k}^\top | a_k \right\} K_\infty A x_k
\end{equation}
according to $a_k$, where the expectations are evaluated with respect to the actual scheduling action $a_k$ and the approximated success rates. This corresponds to a one-step look-ahead controller using $K_\infty$ as terminal costs.

\subsection{DIRA}\label{sec:algorithm}
The combination of the scheduler (Section \ref{scheduler}) and controller designs (Section \ref{sec: controller}) results in the Deep Q-Learning based Iterative Resource Allocation (DIRA) algorithm with time-varying linear quadratic regulation (LQR). 

\begin{algorithm}
	\caption{DIRA with time-varying LQR}
	\label{DQN iterative} 
	\begin{algorithmic}[1] 
		\State Initialize the Q-network weights $\theta$ and $\theta_{\text{target}}$.
		\State Initialize replay memory $\mathcal{R}$ to size $G$.
		\For {the entire duration}
		\State Select action $a_k$ as described in section \ref{scheduler} with exploration parameter $\varepsilon$.
		\State Execute $a_k$ to obtain reward $r_k$ and state $s_{k+1}$.		
		\State Store selection history in $\mathcal{R}$, by associating $r_k$ and $s_{k+1}$ to each intermediate state of step 4.
		\For {each intermediate state}
		\State Sample a random minibatch of $N$ transitions ($s_k,a_k,r_k,s_{k+1}$) from $\mathcal{R}$.
		\State Set $y_j =  r_j + \gamma \max_{a'} Q(s_{j+1},a';\theta_{\text{target}})$
		\State Gradient descent step on $(y_j - Q(s_j,a_j;\theta))^2$.
		\EndFor
		\State $\theta_{\text{target}} = (1-\tau)\theta_{\text{target}} + \tau \theta $.
		\State Every $c$ steps approximate $K_\infty$.
		\EndFor
	\end{algorithmic}
\end{algorithm}

At each time-step $k$ an action is selected in an $\epsilon$-greedy manner. Specifically, we pick a random action $a_k$ with probability $\epsilon$, and we pick a greedy action for all intermediate states as in \eqref{eq: greedy} with probability $1-\varepsilon$. During training, the exploration parameter $\varepsilon$ is decreased to transition from exploration to exploitation.
In step 8, the targets for the Q-Network are computed using a target network with weights $\theta_{target}$. In step 12, the weights $\theta_{target}$ are updated to slowly track the values of $\theta$. This technique results in less variation of the target values, which improves learning, see \cite{mnih2015human}. In step 13, we approximate $K_\infty$ using samples of $\Delta_k$ from the last $D$ time-steps. Updating the control policy at every time-step increases the computational effort since the Riccati equation has to be solved accurately at every time-step. Additionally, updating the control policy frequently induces non-stationarity into the environment, which makes learning difficult. This is similar to the reason why target networks are used. Therefore, we update the control policy only every $c$ steps. 

\begin{rem}
	Acknowledgment of successful transmissions are only necessary in the learning phase. Thus, a converged policy can be used without any communication overhead.
\end{rem}


\section{Numerical results}\label{sec: simulation}
\textbf{Experimental set-up}: We conducted three sets of experiments for a varying number of subsystems and resources. Specifically, we evaluate the scaling of our algorithm by considering the pairs $(N=8, M=6)$, $(N=12, M=9)$, $(N=16, M=12)$.
The systems are generated using random second order subsystems, which are coupled weakly according to a random graph. Regarding stability, we generated 50\% of the subsystems as open-loop stable and 50\% as open-loop unstable, with at least one eigenvalue in range (1,1.5). Additionally, the systems are selected such that the optimal loss per subsystem equals approximately 1. For communication, we consider 
two state Markov models known as Gilbert-Elliot models. We consider two channel types with average error probability $p_1 = 0.99$ and $p_2 = 0.93$.
In every experiment $1/3$ of the channels are of type 1 and $2/3$ are of type 2. 

We would like to highlight, that in all our experiments a uniformly random scheduling policy results in an unstable system. On the other hand, a slightly more ``clever" random policy, which assigns channels randomly according to the degree of stability of each subsystem, is able to stabilize the system in each experiment. In the following we refer to this policy as the ``Random Agent''.

\textbf{Algorithm hyper-parameters:}
In all experiments, the Q-Network is parameterized by a single hidden layer neural network, which is trained using the optimizer ADAM with learning rate $e^{-6}$, see \cite{kingma2014adam}. The training phase is implemented episodically. Specifically, the agent interacts with the system for 75 epochs of horizon $T=500$, where the system state is reset after each epoch. For the three sets of experiments we vary the following parameters: We used $(2048, 4096, 6144)$ for the number of rectifier units in the hidden layer of the Q-Network; we initialized $\varepsilon$ to $1$ and attenuated it to $0.001$ at attenuation rates $(0.99995, 0.99997, 0.99999)$; we used $(75000, 100000, 125000)$ for the replay memory size $G$. In all experiments we used: $\gamma = 0.95$; minibatch size $40$; $\tau = 0.005$; $c=T$; $D=4T$.

\begin{figure}[t]
	\centering
	\includegraphics[width=\linewidth]{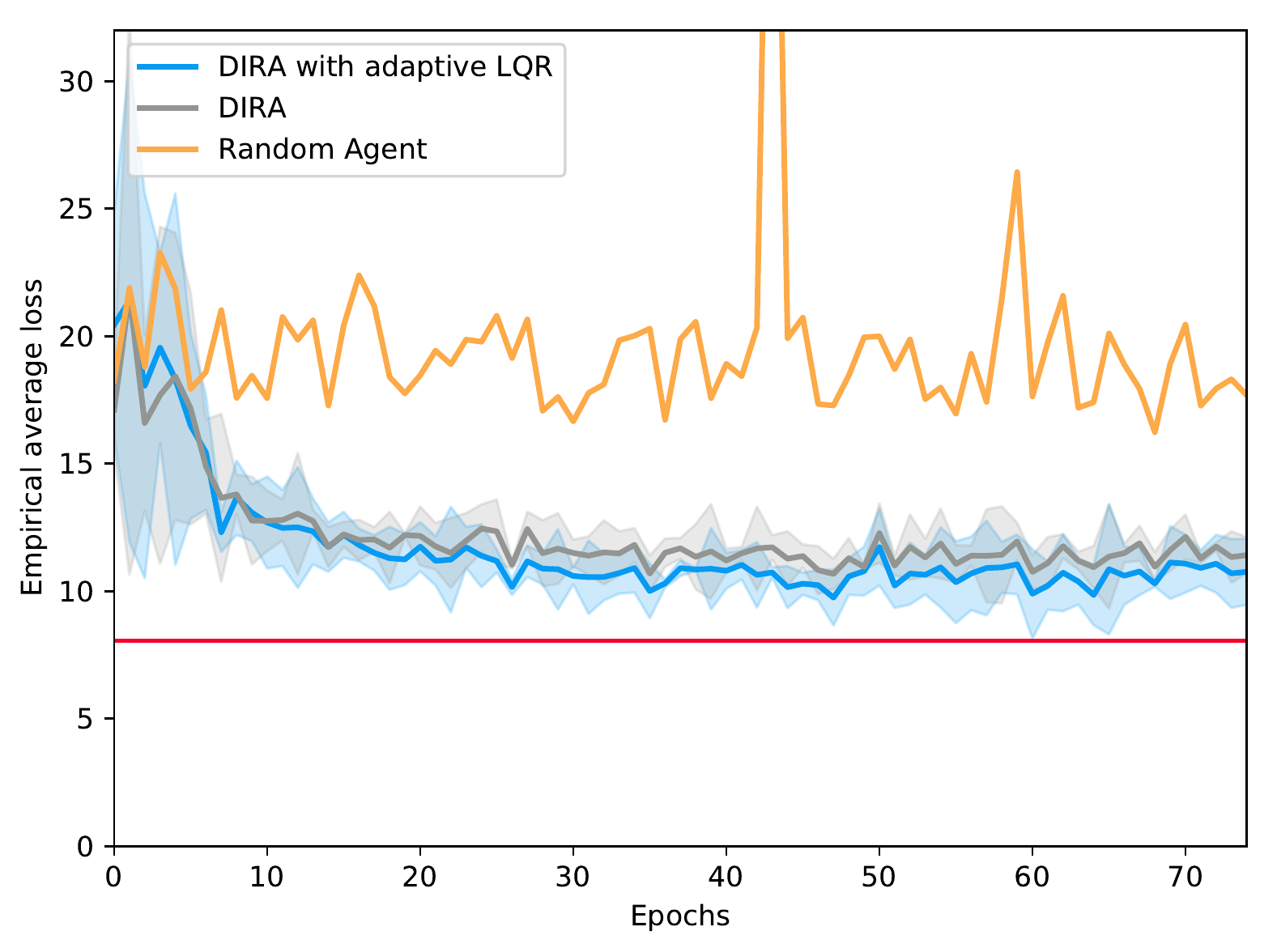}
	\vspace{-0.75cm}
	\caption{Empirical average control loss averaged over 15 training runs for a NCS with $N = 8$ actuators and $M=6$ channels. As a baseline, the red line corresponds to the LQR control loss (no network).}
	\label{fig: learning progress}
\end{figure}

Fig. \ref{fig: learning progress} shows the learning progress of our iterative agent DIRA for ($N$=8, $M$=6) averaged over 15 Monte Carlo runs. For illustration, we compare DIRA with adaptive LQR to DIRA, where $\E\{\delta_k^i\}$ is estimated a priory with samples generated by the Random Agent, and to the Random Agent. For the learning agents, we also display two standard deviations as shaded areas around the mean loss. DIRA with adaptive LQR improves upon the initial random policy and improves further upon the initial estimate of $K_\infty$.  We observed empirically that after convergence of DIRA it is useful to increase $D$ to speed up the convergence of $K_\infty$.
Our algorithm finds a policy, which achieves a control loss of approximately $10.54$ per stage, while the optimal cost per stage under perfect communication is $8.05$. This is significant, especially when taking into account that the neural network architecture is simple.

\begin{figure}[t]
	\centering
	\includegraphics[width=\linewidth]{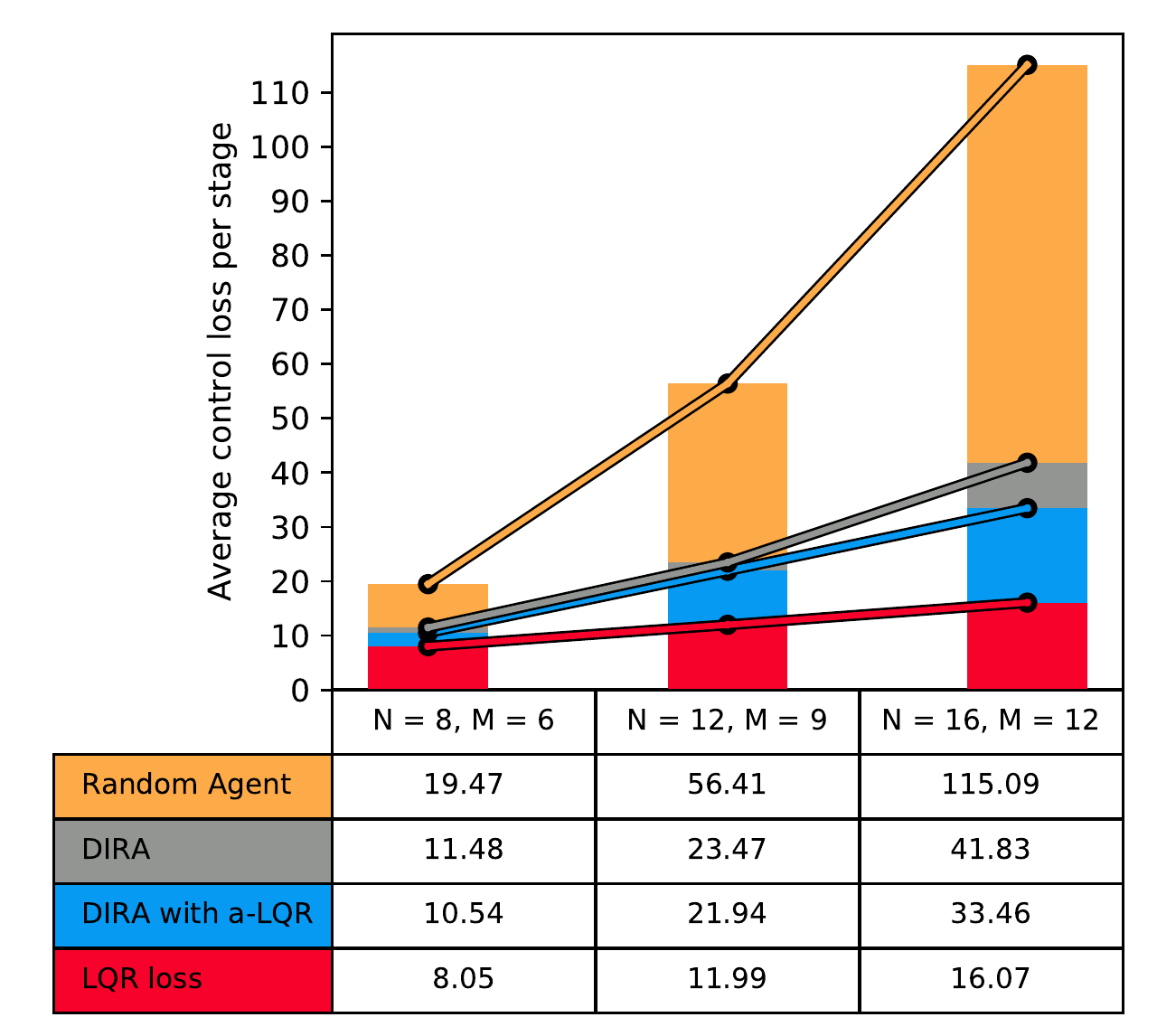}
	\vspace{-0.5cm}
	\caption{Empirical per stage control losses obtained by Monte Carlo averaging using final/trained policies.}
	\label{fig: scale_plot}
\end{figure}

Fig. \ref{fig: scale_plot} compares the training results for all three experiments. We observe that DIRA is able to achieve good performance for the set-ups examined. Recall that in our three experiments the decision space has a size of $N^M$ which is approximately $(2.6 \times 10^5, 5.2 \times 10^9, 2.8 \times 10^{14})$, respectively.
DIRA is able to find good policies in these large decision spaces.
\section{Conclusion}\label{sec:conclusion}
We presented DIRA, an iterative deep RL based resource allocation algorithm for control-aware scheduling in NCS. Our simulations showed that our co-design solution is scalable to large decision spaces. In the future we plan to consider state estimation, scheduling of sensor-controller links as well as the controller-actuator side and time-varying resources. Finally, we are also working towards a theoretical stability result. 

\begin{ack}
The authors would like to thank the Paderborn Center for Parallel Computing (PC$^2$) for granting access to their computational facilities. 
\end{ack}

\bibliography{ifacconf}   
         
\end{document}